\documentclass{article}
\usepackage[a4paper, total={6in, 9in}]{geometry}
\usepackage[utf8]{inputenc}
\usepackage{graphicx}
\usepackage{amsmath}
\usepackage{amssymb}
\usepackage{breqn}
\usepackage{caption}
\usepackage{subcaption}
\usepackage{mhchem}
\usepackage{physics}
\usepackage{multicol}
\usepackage{blindtext}
\normalsize
\title{Channel Capacity of Starch and Glucose Molecular Communications in the Small Intestine Digestive Tract}
\author{Dixon Vimalajeewa and Sasitharan Balasubramaniam }
\date{}

\begin{document}

\maketitle

\section*{Abstract}
The emerging field of Molecular Communication (MC) aims to characterize biological-based signaling environments through information that are encoded into molecules. Since the birth of this field, a number of different applications and biological systems have been characterized using MC theory. This study proposes a new application and direction for MC, focusing on the digestive system, where we characterize and model the starch and glucose propagation along the small intestine (SI). Based on the  advection-diffusion and reaction mechanisms, we define a channel capacity for the small intestine digestive tract that is dependent on the starch to glucose conversion, velocity flow within the tract, viscosity of the digest product, and length of the tract and position of the receivers for glucose absorption. The numerical results from the derived channel capacity model  shows that the SI digestive capacity depends both on physiological factors of the digestive system and type of consumed food, where the digestive capacity is greater for shorter gastric emptying time, low viscosity of the digest product and efficient enzyme activity. We believe that our digital MC model of the digestive tract and lead to personalized diet for each individual, which can potentially avoid a number of different diseases (e.g., celiac disease).

{\bf Keywords:} Molecular Communications, Channel Capacity, Advection-diffusion and reaction system, Small Intestine digestion dynamics.

\section{Introduction}
 
Molecular communications (MC) has transformed the field of communication engineering from the traditional systems that communicate via electromagnetic waves to a new paradigm where information are encoded and transported through molecules. The governing principle of MC is a new perspective of characterizing and modeling biological communication systems that utilize molecules. The uniqueness of MC is the analysis of the communication behavior of the system, which are largely derived from conventional communication systems. These analysis includes determining the capacity of molecule propagation dynamics within the medium channel or the quantity of noise emanating from bio-chemical reactions that impact on the information transport \cite{33}. 

There are a number of different MC systems that have been investigated by the community and they differ from the types of cells, molecules, as well as range of communications. For example,  short-range communication in MC includes calcium-based signaling where information are encoded into ions, while medium range is the use of bacteria to transport DNA encoded information. An interesting direction that has been taken by the community is long-range MC, where information molecules are transported between different organs. A good example is MC that takes place within the circulatory system, and an application that researchers have focused on is drug delivery. This also facilitates new approaches to using engineered nano-bio science to control the communication processes and provide innovative solutions to overcome different challenges and enhance performance in various application domains. For example, site-specific targeted drug delivery to maximize the efficacy of drug molecules while minimizing the potential side effects in healthcare \cite{37, 38} and neuron communications to characterize different information flows in the bran \cite{7}.  The community has embraced the concept of MC as a typical communication system that can be connected as part of the Internet of Things. This is known as the Internet of Bio-Nano Things, where molecular communication nano-networks can transmit information to the Cyber-Internet \cite{40, 41}. 

%In doing that, MC uses the theories from conventional communication systems to model molecule propagation dynamics along a particular medium such as within or between cells. Those models are then used to design computational tools to analyze communication mechanisms in MC systems . 
%This in turn increases the potential of exploring complex biological communications that occur in natural biological systems. 
  
\begin{figure}[!t]
    \centering
    %\vspace{-1.5cm}
    \includegraphics[width = \textwidth]{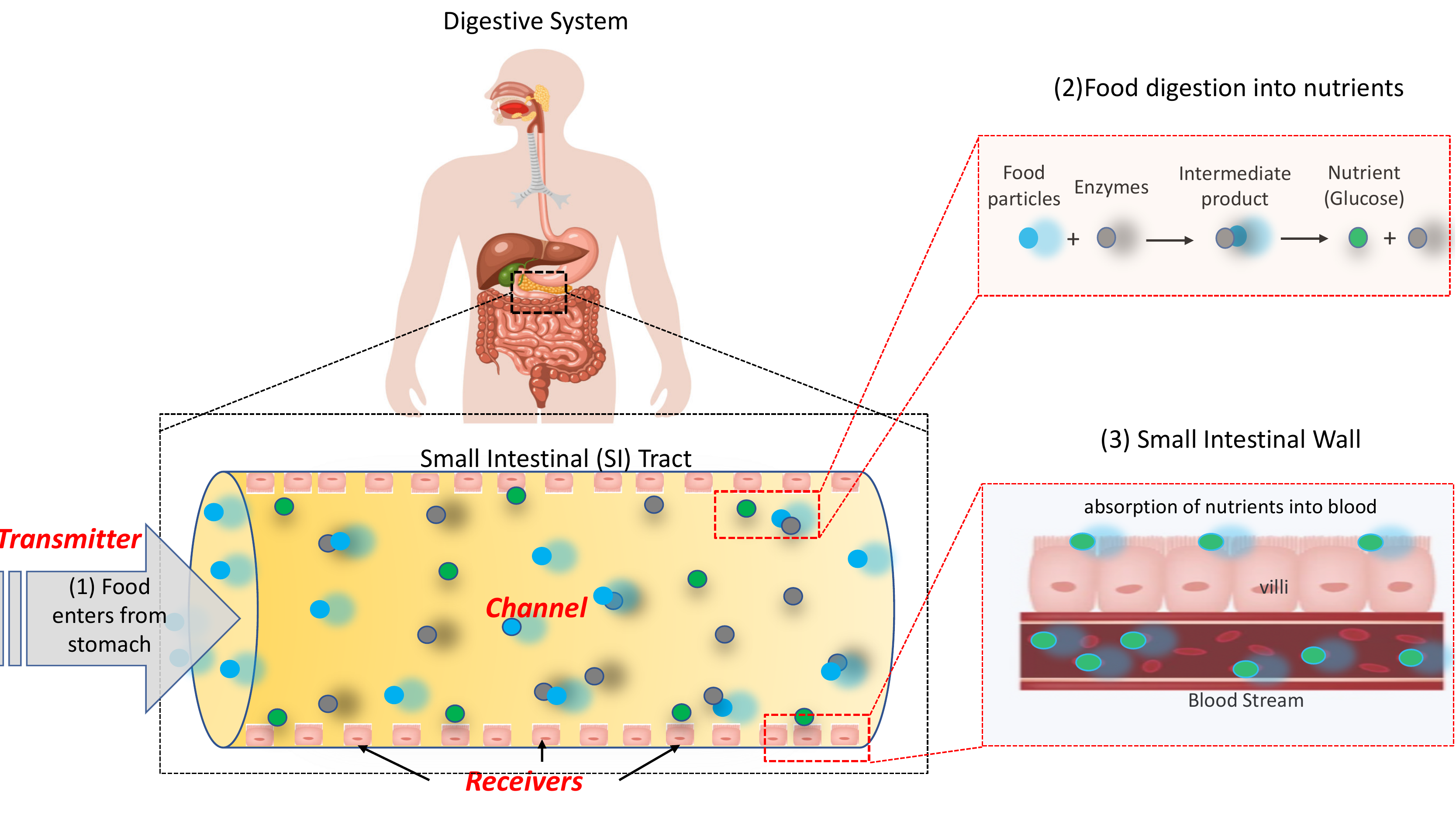}
    %\vspace{-2.5cm}
    \caption{An overview of the main stages of the digestion process and mapping them to the components of a molecular communication system: food particles enter from the stomach (\emph{transmitter}) interact with enzymes while traveling along the small intestinal (SI) tract (\emph{channel}). Nutrients are produced as a result and then they are absorbed into blood through the SI tract wall (\emph{receiver}). }
    \label{Digest}
\end{figure}

This paper proposes a new type of long-range MC system that links to human health. The study focuses on long-range digestive tract MC to explore digestive system dynamics. While a number of works have been developed for the digestive systems, these works have largely focused on computational models to characterize the system dynamics based on the physio-chemical characteristics and functions \cite{5,6, 19, 20}. Our study, on the other hand, focus on the characterization of the digestive functions and propagation of glucose along the small intestine (SI), where our mapping to the MC paradigm is the channel representation of the molecules that propagate along the SI. Our capacity model considers the dynamics within the SI, as the glucose and starch molecules that propagate and can bind to a number of receivers that is represented as the tissue linings along the SI. The three main stages of the food digestion process within the SI tract are as follows: 1) food content that enters into the SI tract, 2) propagation of starch molecules along the SI tract while converting into glucose, and 3) absorption of glucose into the tissue lining that eventually enters the blood stream. The three stages are illustrated in Figure \ref{Digest} and are mapped to the representative components of a MC system, namely Transmitter (food input), channel (SI tract that facilitates the flow of starch and transformation process into glucose) and receiver (glucose binding along the tissue lining). The key benefit of using the proposed approach compared to the  existing models is that it enables applying concepts in conventional communication theory to explore digestive dynamics. This in turn would enable to expand  characterizing the digestive system dynamics based on different properties of molecules of the consumed food in addition to physicochemical properties.

The study consists of four main steps. First, the study derives a generalized mathematical model to characterize each stage of the digestive system dynamics. This model is used in the second step to derive a set of mathematical models to represent the three stages of the digestion process in terms of carbohydrates digestion within the SI tract. In the third step, the set of mathematical models derived in the previous step is solved analytically by mapping to the MC paradigm to derive an expression to characterize the  channel capacity of the starch and glucose propagation after conversion, as well as the adsorption process. Finally, the study explores the influence of different factors related to the digestive system and consumed food on the channel capacity, linking the efficiency of the glucose conversion process and characterizing the propagation dynamics from a communication perspectives. %The channel capacity is used here because, in communication theory, channel capacity enables characterizing information flow dynamics in a communication system, for a given input signal. Hence, 
The capacity of the MC system proposed in this paper can provide a new analysis tool to understanding the variability of the digestive dynamics within the SI tract. In particular, this can contribute towards the characterization of the  carbohydrates digestion since it represents a major component for the human diet and is the main source of fuel to generate energy required for the body functions. This study assumes that carbohydrates remain intact until it reaches the SI, but in reality, starch digestion starts at the mouth. Moreover, since starch is the main source of carbohydrates, starch is used instead of carbohydrates,  hereafter. Implementing a virtual MC model of the digestive system can lead towards a digital twin of the human gut. Future biosensors that provide vital signals from within the gut can be transmitted to the digital twin, providing us with an accurate view of the SI dynamics based on the diet that is consumed

%Implementing a virtual patient is an example application where bio-inspired communication concepts are utilized in which various treatment models are created to standardize clinical knowledge and medical data. Those models are then used to evaluate treatment efficacy and toxicity, aiming to improve patient survival and quality of life.

The reminder of the paper is organized as follows: Section \ref{sec-2} gives an overview of the system model, and the results derived from the system model are presented in section \ref{sec-3}. The potentials of the study in expanding the digestive system dynamics as a MC system are discussed in section \ref{sec-4} and this is followed by section \ref{sec-5} that concludes the study.

%%###################################################
\section{System Model}\label{sec-2}
During the digestive process, food undergoes several intermediate stages prior to absorption of nutrients  into the blood stream which in turn will distribute the molecules throughout the body to be used as energy to support various body functions. The system model proposed below considers the digestive system as a molecular communication system within the SI tract by considering the advection-diffusion and reaction process. That is, functionality of the main components of the digestive system, namely the stomach, SI tract, SI wall are mapped to the functionality of the MC systems components, namely the transmitter, channel and receiver, which is explained below (illustrated in Figure \ref{Digest}).  

%\begin{figure}[!t]
%    \centering
%    \vspace{-1.0cm}
%    \includegraphics[width = .6\textwidth]{Fig-New/Fig12.pdf}
%    \vspace{-2.0cm}
%    \caption{System model: mapping the main stages of the digestive system illustrated in Figure \ref{Digest} into a molecular communication system. }
%    \label{Digest2}
%\end{figure}

\begin{itemize}
    \item [(1)] {\bf Transmitter} is  the stomach as it releases digested food into the SI tract, which acts as a reservoir for gastric residue (note that mechanically broken down food in the mouth is mixed with gastric juice). %The rate gastric content is released into the SI tract (known as gastric emptying rate) influences on the digestion and abundance of nutrient in the MC channel.  
    
    \item [(2)] {\bf Channel} is the SI tract as it provides a medium to convert food particles into absorbable glucose (i.e., digestion) while moving along the SI tract. The contraction and relaxation of SI muscles generate convective flow of digestion while the concentration gradient of nutrient molecules, due to enzyme reaction on digest along the SI tract, creates a diffusive flow of nutrient molecules. Thus, this process is similar to the functionality of an advection-diffusion and reaction based  MC channel.

    \item [(3)] {\bf Receiver} is the tissues linings along the SI wall which absorb the glucose molecules and diffuses into the circulatory system. The SI wall is made of {\it villi}, which are tissue projections (or folds) and has the role of increasing the surface area to absorb glucose effectively.
\end{itemize}

The study assumes carbohydrates remain intact until it reaches the SI tract though carbohydrates digestion that starts at the mouth itself, but that is approximately 5\%. %Since starch is the main source of carbohydrates, hereafter, starch is used instead of carbohydrates. 
The system model discussed below first presents a model that characterizes the flow transport along the SI tract (section \ref{A}). This is followed by a system of equations to represent the digestion of starch into glucose (section \ref{B}). The analytical solutions derived for the system of equations are then used to derive an expression for the channel impulse response (section \ref{C}). Finally, the channel impulse response is used to compute the channel capacity (section \ref{D}) which is used in the next section \ref{sec-3} to explore the digestive system dynamics.

\subsection{Food Transport Model}\label{A}
This model primarily aims at the digestion and absorption of glucose take place within the SI tract. The mechanical force generated by the SI tract assists the digest product to move along the SI tract which in turn creates am advection flow of average velocity $u$. Along the propagation path, the enzyme reaction on digest product creates absorbable nutrients at a rate $C_P(x,t)$ at an arbitrary point $x$ and time $t$ and this product  diffuses  through the SI wall  into the blood stream at a rate $K (= \frac{2}{d} K_a f)$, where $d$ is the radius of the SI and $f$ is the SI wall surface area increasing factor due to villi folds. This result in a diffusion flow of nutrient molecules with a diffusion coefficient $D$. The (\ref{nu-6}) accounts these mechanisms and expresses the changes in the nutrient concentration ($C(x,t)$) within the SI tract for time $t$, and is represented as follows 

\begin{equation}\label{nu-6}
    \frac{\partial C(x,t)}{\partial t} + u \frac{\partial C(x,t)}{\partial x} = D \frac{\partial^2 C(x,t)}{\partial x^2} - C_P(x,t) - KC(x,t).  
\end{equation}
Based on the Michaelis-Menten-kinetics related to the enzyme reaction on the digest product \cite{17}, the glucose production rate, $C_P(x,t) = V_{m} \frac{C(x,t)}{K_{m} + C(x,t)} (s^{-1})$, where $V_m (Ms^{-1})$ is the maximum reaction rate achieved at the maximum saturating concentration and the half saturation concentration $K_{max} (M)$. The mass transfer coefficient, $K_a = 1.62 \left(\frac{u D^2}{Ld}\right)^{\frac{1}{3}} (ms^{-1})$; the diffusion coefficient is $D = \frac{K_BT}{6\pi \mu r_m} (m^2 s^{-1})$, $K_B (m^2kg s^{-1}K^{-1})$ is the Boltzmann constant, $T (K)$ is the temperature and $r_m (nm)$ is the radius of the nutrient molecules and $L (m)$ and $d (cm)$ are the length and diameter of the SI tract, respectively  (detailed derivation for (\ref{nu-6}), $C_P(x,t)$, and $K_a$ is given in Appendix).

%%########################################################
\subsection{Starch Digestion Model}\label{B}
Starch digestion into glucose and subsequently absorption into the blood stream in the SI tract is formulated as a system of differential equations (\ref{eq-1}) - (\ref{eq-3}) by using (\ref{nu-6}). The expressions derived for the $C_{P}(x,t)$ and $K$ are utilized in deriving these models and they are given in the Appendix.

 \begin{eqnarray}\label{di-eqn-new}
\dv{C_{st}(t)}{t} &=& -\gamma C_{st}(t), \label{eq-1}\\
\frac{\partial C_s(x,t)}{\partial t} &=& D_s \frac{\partial^2 C_s(x,t)}{\partial x^2} - u \frac{\partial C_s(x,t)}{\partial x} - V_m\frac{C_s(x,t)}{K_m+C_s(x,t)} + \gamma C_{st}(t), \label{eq-2} \\
\frac{\partial C_g(x,t)}{\partial t} &=& D_g \frac{\partial^2 C_g(x,t)}{\partial x^2} - u \frac{\partial C_g(x,t)}{\partial x} + V_m\frac{C_s(x,t)}{K_m+C_s(x,t)} - K C_{g}(x,t), \label{eq-3}
\end{eqnarray}
with boundary conditions $C_{st}(t =0) =C_0, C_s(x = 0,t =0) = C_g(x=0, t = 0) = 0$, where $D_s$ and $D_g$ are the diffusion coefficients of starch and glucose, respectively.

%In response to the stomach releases gastric volume into the SI tract, starch concentration in the stomach ($C_{st}(t)$) decreases at a rate $\gamma C_{st}$ (see equation \ref{eq-1}, where $\gamma = \log(2)/st$ and $st$ is the half-gastric emptying time) and in turn the starch concentration in the SI tract ($C_s(x,t)$) increases at the same rate. 
The starch concentration in the stomach ($C_{st}(t)$) decreases at a rate $\gamma C_{st}$ as the gastric content in the stomach is released into the SI tract (see (\ref{eq-1}), where $\gamma = \log(2)/st$ and $st$ is the half-gastric emptying time). This results in an increase in the starch concentration in the SI tract ($C_s(x,t)$) along distance $x$ at the same rate. While starch molecules travel along the SI tract under the advection-diffusion mechanism, the reaction of the digestive enzymes hydrolyses starch which is subsequently converted into glucose by the brush broader enzymes \cite{30}. As a result, the starch concentration $C_s(x,t)$ decreases along the SI tract at a rate  $C_P(x,t) = V_m\frac{C_s}{K_m+C_s}$ (see  (\ref{eq-2})) and the glucose concentration ($C_g(x,t)$) increases at the same rate. In the meantime, as given in  (\ref{eq-3})), the glucose concentration $C_g(x,t)$ decreases at a rate $KC_g(x,t)$ as the SI wall absorbs glucose into blood. %while they travel along the SI tract under advection-diffusion mechanisms . 

%%##############################################################
\subsection{Channel Impulse Response}\label{C}
In molecular communication theory, the channel impulse response reflects the expected number of molecules at the receiver, given that a certain number of messenger molecules has been released by the transmitter \cite{1, 24}. The channel impulse response for the MC system corresponding to the digestive system is derived by solving (\ref{eq-1})-(\ref{eq-3}) analytically.

%The half-saturation concentration $K_m $ used in (\ref{eq-1})-(\ref{eq-3}) reflects the relative ability of enzymes to use low levels of starch while $V_m$ stands for maximum growth rate of glucose concentration {\bf**odd sentence}. 

A large $K_m$ value (i.e., $K_m >> C_s$) implies that maximum growth rate of glucose concentration $V_m$ is mostly controlled by the starch concentration. On the other hand, when $K_m << C_s$, $V_m$ is maximal and independent from starch concentration, which means that all enzymes have bound to the starch molecules \cite{31}. So, in reality, it can be assumed that there is an excess  amount of enzymes compared to the starch molecules within the SI tract. Moreover, glucose dynamics within the SI tract also depend on the starch concentration. Therefore, it is assumed that $K_m >> C_s$ for solving the systems of equations (\ref{eq-1})-(\ref{eq-3}) to get the glucose concentration $C_g(x,t)$ at an arbitrary point $x$ and time $t$ within the SI tract, and is represented as follows 

    \begin{equation}\label{eq-3-soln}
        C_g(x,t) = \frac{C_0\gamma V_mG}{(V_m - KK_m)(K- \gamma)}\left[ \frac{1}{\sqrt{4\pi D_g t}} e^{\left(\frac{-(x-ut)^2}{4D_gt} - Kt\right) } - \frac{1}{\sqrt{4\pi D_s t}} e^{\left( \frac{-(x-ut)^2}{4D_st}  -\frac{V_m}{K_m}t \right)} + e^{-\gamma t} \right].
    \end{equation}
    
The glucose concentration $C_{G}(t)$ within the SI tract at time $t$ is computed  by integrating (\ref{eq-3-soln}) over the SI tract distance ($L$) and can be expressed as

\begin{eqnarray}\label{eq-3-soln-1}
    %C_{G}(t) &=& \frac{C_0 \gamma V_m}{(V_m - KK_m)(K-\gamma)} \left[ \frac{1}{2}\left( e^{-Kt} - e^{-\frac{V_m}{K_m}t}\right)  \left( erf \left(\frac{d-ut}{2\sqrt{Dt}}\right) + erf \left(\frac{ut}{2\sqrt{Dt}}\right) \right)+ de^{-\gamma t} \right]
    C_{G}(t) = \frac{C_0 \gamma V_m}{(V_m - KK_m)(K-\gamma)} \left[ \frac{L}{2} e^{-Kt} \left( erf \left(\frac{L-ut}{2\sqrt{D_gt}}\right) + erf \left(\frac{ut}{2\sqrt{D_gt}}\right) \right)\right] \nonumber \\
    -\frac{C_0 \gamma V_m}{(V_m - KK_m)(K-\gamma)} \left[ \frac{L}{2}e^{-\frac{V_m}{K_m}t} \left( erf \left(\frac{L-ut}{2\sqrt{D_gt}}\right) + erf \left(\frac{ut}{2\sqrt{D_gt}}\right) \right) +
    le^{-\gamma t} \right].
\end{eqnarray}

Since the SI wall is an absorbing receiver \cite{3}, the expected number of glucose molecules absorbed into the blood stream through the SI wall can be expressed as 

\begin{equation}\label{eq-3-soln-prob}
    C_{SI \rightarrow B}(t) = C_0 - C_{G}(t) = C_0 f(t|L),
\end{equation}
where  $f(t|L) = 1 -  \frac{ \gamma V_m}{(V_m - KK_m)(K-\gamma)} \left[ \frac{L}{2} e^{-Kt} \left( erf \left(\frac{L-ut}{2\sqrt{D_gt}}\right) + erf \left(\frac{ut}{2\sqrt{D_gt}}\right) \right)\right] 
   + \frac{ \gamma V_m}{(V_m - KK_m)(K-\gamma)}  \\ \left[ \frac{L}{2}e^{-\frac{V_m}{K_m}t} \left( erf \left(\frac{L-ut}{2\sqrt{D_gt}}\right) + erf \left(\frac{ut}{2\sqrt{D_gt}}\right) \right) +
    xe^{-\gamma t} \right]$.

Given that the transmitter has released $N$ number of starch molecules, the expected number of molecules at the receiver at time $t$ can be expressed as $\mu(t) = NP(x,t)$, assuming molecules move independently, where $P(x,t)$ is the probability of the activation of receptor molecules at the receiver. Therefore, the channel impulse response ($h(t|x)$, where $x$ is the distance) of the MC channel considered above can be represented based on (\ref{eq-3-soln-prob}) as:

\begin{multline}
%P(x,t) = 1 - \frac{V_m \gamma}{(V_m - KK_m)(K-\gamma)} \left[ \frac{x}{2}\left( e^{-Kt} - e^{-\frac{V_m}{K_m}t}\right)  \left( erf \left(\frac{x-ut}{2\sqrt{Dt}}\right) + erf \left(\frac{ut}{2\sqrt{Dt}}\right) \right)+ xe^{-\gamma t} \right]
h(t|x) = 1 -  \frac{ \gamma V_m}{(V_m - KK_m)(K-\gamma)} \left[ \frac{x}{2} e^{-Kt} \left( erf \left(\frac{x-ut}{2\sqrt{D_gt}}\right) + erf \left(\frac{ut}{2\sqrt{D_gt}}\right) \right)\right] 
   + \\ \frac{ \gamma V_m}{(V_m - KK_m)(K-\gamma)}
   \left[ \frac{x}{2}e^{-\frac{V_m}{K_m}t} \left( erf \left(\frac{x-ut}{2\sqrt{D_gt}}\right) + erf \left(\frac{ut}{2\sqrt{D_gt}}\right) \right) +
    xe^{-\gamma t} \right].
\end{multline}

In terms of probability theory, $h(t|x)$ represents a conditional probability, meaning that molecules receiving (i.e., output) probability at time $t$, given that transmitted molecules (i.e., input) have traveled $x$ distance from the transmitter. However, the receiver will not detect the same number of molecules as transmitted within a certain time interval due to interference such as channel noise, which can be Brownian of residual noise. Therefore, these factors are considered when we determine the channel capacity in the next section. 
%%===================================================================
%%===================================================================
\subsection{Small Intestine Channel Capacity}\label{D}
We assume in our digestive tract operation scenario that the release of the digest product from the stomach and detection of glucose molecules by the SI wall events occur over a discrete time period  $T$ such that $T = nt_W$, where $n$ is the number of time windows and $t_W$ is the time window duration. The molecules released into the MC channel have a stochastic movement, where they follow different trajectories to reach the receiver. This results in a delay in receiving molecules by the receiver, increasing the signal noise which in turn limits the channel capacity ($C$). The binary MC channel based approach in \cite{23} is used to formulate the small intestine channel capacity $C$, but there are studies that have used different approaches such as free molecule diffusion to compute the channel capacity in MC systems \cite{39}. A binary channel means that the transmitter sends either a $0$ or $1$ digits to the receiver (in our case the molecule transmission represents $1$, but a $0$ is a silent transmission). Please see \cite{23, 24} for more information about the binary channel-based capacity analysis and the use of information theory in computing capacity \cite{35}. 

Suppose the transmitter releases $N$ number of molecules to signify that it sends $1$ bit and releases of no molecules implies that sends $0$ bit. Since molecules move freely (i.e., independently) along the channel, the number of molecules that reach the  receiver follows a Binomial distribution  $\mathcal{B}(N,P(x,t_W))$. Hence, the total number of molecules  detected by the receiver over a time window $t_W$ (say $\hat{S}(t_W)$) can be written as $\hat{S}(t_W) = S(t_W) + n_R(t_W) + n_B(t_W) $, where
%({\bf have to make sure T or t for $\hat{S}$})
\begin{itemize}
    \item[-] {\bf  $S(t_W)$} is the signal generated in response to the release of $N$ molecules by the transmitter in time window $t_W$. Given that $b_c$ is the prior transmission probability of $N$ molecules within the time window $t_W$, $S(t_W)$ has a Binomial distribution represented as 
    \begin{equation}
        S(t_W) \sim \mathcal{B}\left(N, P(x,t_W)b_c\right),
    \end{equation}
    
    \item[-] {\bf  $n_R(t_W)$} is the residual noise (or inter-symbol interference (ISI) \cite{36}) that occurs due to  delay in receiving molecules within the same time window that they were transmitted. This occurs when certain molecules transmitted in time slots $0,1, \cdots, t_{W-1}$ are being received in the time window $t_W$. Suppose $b_k$ is the prior probability of the delay in receiving molecules within the same time window that they were transmitted, then the probability of receiving molecules in the $n^{th}$ time window which were emitted in the $i^{th}$ time window is $P_{R} = [P(x,(n-i)t_W) - P(x, (n-(i-1))t_W)]b_k$ \cite{8, 9}. This leads to residual noise $n_R(t_W)$ having the probability distribution represented as
    \begin{equation}
        n_R(t_W) \sim  N\sum_{i=1}^{n-1}\mathcal{B} (N, P_{R}), 
    \end{equation}
    where $P_R = (P(x,(n-i+1)t_W) - P(x, (n-i)t_W))b_k $.
    \item [-] {\bf $n_B(t_W)$} is the Brownian (or background) noise that occurs due to random movement of the transmitted molecules and assumed to have an additive white Gaussian distribution  with mean zero and standard deviation $\sigma$ and is represented as
    \begin{equation}
        n_B(t_W) \sim \mathcal{N}(0, \sigma^2).
    \end{equation}
\end{itemize}

Based on statistics, for large $N$ value, $NP(x,t_W)$ is not close to zero so  this facilitates us to make a binomial-normal approximation. Suppose a random variable $Z$ has a Binomial distribution with success probability $p$ over $n$ number of trials, this results in $Z\sim \mathcal{B}(n, p)$. For large $n$, $Z$'s probability distribution is close to a Gaussian distribution with mean $\mu = np$ and standard deviation $\sigma = np(1-p)$ (denoted as $Z \sim \mathcal{N} (np, np(1-p))$. 

Upon the release of $N$ number of molecules by the transmitter to signify sending  $1$ bit, the receiving the transmitted  $1$ bit is decided by using a threshold value ($Ep$) represented as:
\begin{align}
\hat{S}(t_W) = \left\{ \begin{array}{cc} 
                0 & \hspace{5mm} \hat{S_T}(t_W) < Ep, \\
                1 & \hspace{5mm} \hat{S_T}(t_W) \geq Ep. \\
                \end{array} \right.
\end{align}

Since both the input and output has values $0$ and $1$,  the sample space is $ s =  \{00, 01, 10, 11\}$ and their probabilities can be derived as follows:

\begin{itemize}
    \item[-] {\bf $P(\hat{S} = 0|S = 0)$} represents that there is no emission of molecules from the transmitter, but the receiver may still receive some delayed molecules from previous time slots. That is, $S(t)=0$, but  $\hat{S} = n_R(t_W), n_B(t_W) \neq 0 $ and $n_R(t_W)+ n_B(t_W) < Ep$, and this leads to the following
    \begin{equation}
        P(\hat{S} = 0| S = 0) = P(\hat{S_T} < Ep) = P(n_R + n_B < Ep) = P(z < \frac{Ep - \mu_1}{\sigma_1}), \nonumber
    \end{equation}
    where $\mu_1 = N\sum_{i=1}^{n-1}P_{R} $ and $\sigma_1^2 = N\sum_{i=1}^{n-1} P_{R}(1-P_{R}) + \sigma^2$.
    %Given that there is no release of molecules, if not the event $\hat{S} = 0| S = 0$, 
    Another case is $ \hat{S} = 1| S = 0$ and this can occur with the following probability,
    \begin{equation}
        P(\hat{S} = 1| S = 0) = P(\hat{S_T} \geq EP) = P(n_R + n_B \geq Ep) = 1-  P(n_R + n_B < Ep). \nonumber
    \end{equation}
    
    \item[-] {\bf $P(\hat{S} = 0| S = 1)$} means that there is an emission of molecules from the transmitter, but the receiver may still not be able to absorb a sufficient amount of molecules to be considered as $1$ bit  due to the delay in receiving molecules. That is,  $S(t_W), n_R(t_W)$ and $n_B(t_W)  \neq 0$, but  $\hat{S} = S(t_W) + n_R(t_W)+ n_B(t_W) < Ep$ and this is represented as follows
    \begin{equation}
        P(\hat{S} = 0| S = 1) = P(\hat{S_T} < Ep) = P(S + n_R + n_B < Ep) = P(z < \frac{Ep - \mu_2}{\sigma_2}), \nonumber
    \end{equation}
    where $\mu_2 = NP(x,t_W)b_c + N\sum_{i=1}^{n-1}P_{R}$ and $\sigma_2^2 = NP(x,t_W)b_c(1-P(x,t_W)b_c)+ N\sum_{i=1}^{n-1}P_{R}(1 - P_{R}) + \sigma^2$.
    Similar to the previous step, $\hat{S} = 1| S = 1$ is the other possible event if the event $\hat{S} = 0| S = 1$ does not occur and its probability is represented as follows 
    \begin{equation}
        P(\hat{S} = 1| S = 1) = P(\hat{S_T} \geq Ep) = P(S + n_R + n_B \geq Ep) = 1-  P(S + n_R + n_B < Ep). \nonumber
    \end{equation}
\end{itemize}

These probabilities are used to compute the mutual information, $MI(S,\hat{S})$ in (\ref{MI}) and this quantifies the information flow carried through the channel over time window $t_W$, given that $N$ number of molecules are released by the transmitter, and the this is represented as follows 

\begin{equation}\label{MI}
    I(S,\hat{S}) = \sum_{S = \{0,1\}} \sum_{\hat{S} = \{0,1\}} P(S,\hat{S})\log_2 \left(\frac{P(\hat{S}|S)}{P(\hat{S})}\right).
\end{equation}

Finally, the small intestine capacity of the MC channel is computed as:

\begin{equation}\label{capa}
     C = \underset{Ep}{max} I(S,\hat{S}).
\end{equation}

%%======================================================

\section{Numerical Results}\label{sec-3}
%%======================================================
The model derived in section \ref{sec-2} is used here to explore the impact of physiological factors associated with the digestive system and the type of consumed food on the small intestine digestive capacity in the context of carbohydrate digestion. Table \ref{tab-1} lists all the parameters that is used for our simulations, which was built using the Python programming language.

\begin{table}[t!]
\centering
\caption{Model Parameters Used for Simulations.}
\begin{tabular}{lll}
\hline
Parameter & Symbol & Value  \\
\hline
%{\bf Digestive System} & \\
\hspace{.25cm} Small intestine length           & $L$ & $ 6.9 m$ \\
\hspace{.25cm} Small intestine diameter         & $d$ & $1.8 cm$ \\
\hspace{.25cm} Viscosity              &$\mu$  & $0.001 - 10 PaS$ \\
\hspace{.25cm} Radius of glucose molecules   & $r_m$ & $ 0.38 nm$\\
\hspace{.25cm} Surface area increase due to fold, vili and microvili 
                                                     & $f$ & $ 12$\\
\hspace{.25cm} Mean velocity.   & $u$ & $ 1.7 \times 10^{-6} m/s $ \\
\hspace{.25cm} Maximum reaction rate  & $V_{m}$& $ 1.95 mM/min $ \\
\hspace{.25cm} Half saturation concentration & $K_{m}$& $ 35 mM$\\
\hspace{.25cm} Half gastric emptying time  & $st$ & $ 1 hour $ \\
\hspace{.25cm} Background noise  & $sig$ (or $\sigma^2$) & $ 150 $ \\
\hspace{.25cm} Time window duration  & $t_W$ & $ 1 hour $ \\
\hspace{.25cm} Number of time windows  & $n$ & $ 10 $ \\
\hline
\end{tabular}\label{tab-1} 
\end{table} 

%%######################################################
\subsection{Starch Digestion}
\begin{figure}[t!]
\centering
\begin{subfigure}{.45\textwidth}
  \centering
  % include second image
  \includegraphics[width=.8\linewidth]{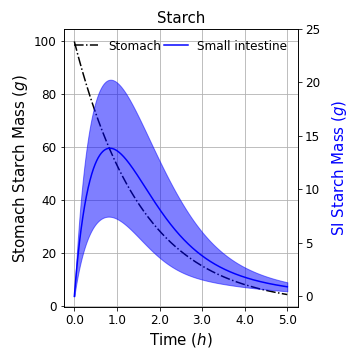}  
  \caption{}
  \label{Res-11}
\end{subfigure}
\begin{subfigure}{.45\textwidth}
  \centering
  % include first image
  \includegraphics[width= .8\linewidth]{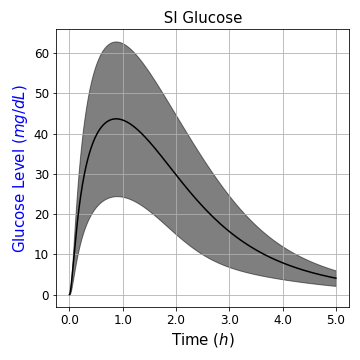}  
  \caption{}
  \label{Res-12}
\end{subfigure}
\caption{Results for $100 g$  of carbohydrates consumed resulting in variations with respect to time for the (a) starch mass in the stomach and the SI tract, and (b) glucose concentration in the SI tract.}
\label{R1}
\vspace{-.75cm}
\end{figure}

Given that $100g$ of carbohydrates is consumed, Figure \ref{R1} show the variability in starch mass in the stomach and SI tract and also the glucose  concentration in the tract over time. As the digest product enters into the SI tract from the stomach, the starch mass is decreasing and Figure \ref{Res-11} shows that approximately $50\%$ of the digest product enters into the SI tract for nearly within an hour and the remaining starch takes around three hours to enter the SI tract. As a result, the starch mass in the SI tract reaches its peak after an hour. This is due to the conversion process of starch into glucose. This results in an increase in glucose concentration and the highest achieved is approximately $42mg/dL$ as depicted in  Figure \ref{Res-12}. The absorption of the produced glucose into the blood stream causes a decrease in the glucose concentration and more than $90\%$ of the produced glucose has been absorbed into the blood stream within three hours. These dynamics could,  however, be varied depending on a number of factors associated with the digestive system (e.g., enzymes within the SI tract) as well as the digest properties (e.g., viscosity). Therefore, the following sections discusses the impact of such parameters on the effective absorption of glucose into blood in terms of variability in the SI channel  capacity.

%%#################################################################
\subsection{Impact of Model Parameters on Mutual Information}
Based on section \ref{sec-2}, the transmission probabilities ($b_c, b_k$), time window size ($t_W$), number of time windows ($n$), threshold ($Ep$) and Brownian noise ($\sigma$) are model parameters that are used to derive the mutual information $MI$, which means that their optimal values have to be selected for accurate computation. Besides, parameters associated with the digestive system such as velocity $u$ and the traveling distance $L$ can also make a significant impact on the mutual information $MI$. %Exploring the impact of those parameters on m $MI$ is essential for computing capacity. 

\subsubsection{Transmission Probabilities ($b_c$ and $b_k$)}
\begin{figure}[t!]
\centering
\begin{subfigure}{.3\textwidth}
  \centering
  % include second image
  \includegraphics[width=1\linewidth]{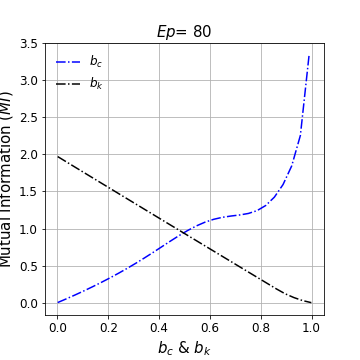} 
  \caption{}
  \label{Res-21}
\end{subfigure}
\begin{subfigure}{.6\textwidth}
  \centering
  % include first image
  \includegraphics[width= 1\linewidth]{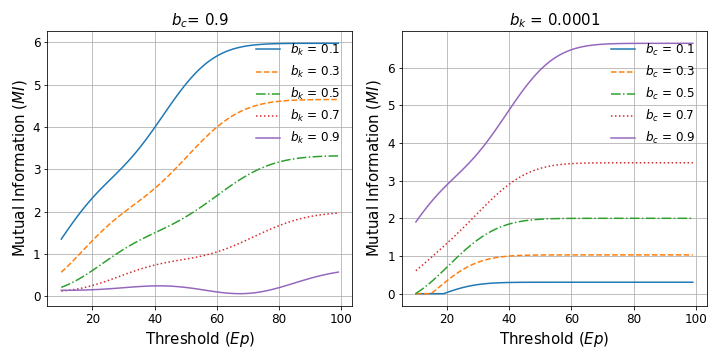}  
  \caption{}
  \label{Res-22}
\end{subfigure}
\caption{Dynamics of the mutual information $MI$ with respect to variations in the (a) transmission and residual transmission probability ($b_c$ \& $b_k$) for a selected threshold, $Ep = 80$ and (b) threshold ($Ep$) when $b_c = 0.9$ and $b_k = 0.0001$.}
\label{R2}
\vspace{-.75cm}
\end{figure}

Figure \ref{Res-21} shows the impact of $b_c$ and $b_k$ on the mutual information $MI$, when other parameters are set as $Ep = 80, t_W =60 min$, and $n = 20$. The greater the $b_c$ value and the lower the $b_k$ value, the larger the mutual information $MI$. Exploring the impact of the $Ep$ threshold on the mutual information $MI$, Figure \ref{Res-22} depicts the change in the mutual information $MI$ with $Ep$ when $b_c =0.9$ and $b_k$ is varying in the range $0-1$ and $b_k =0.0001$ and $b_c$ is varying in the range $0-1$. Both graphs show that the mutual information $MI$ reaches its peak value and then saturates after a certain $Ep$ value, which is higher for smaller $b_k$ values and smaller for higher $b_c$ values. Moreover, the $Ep$ threshold value that maximizes the mutual information $MI$ (i.e., capacity) increases with smaller $b_k$ and higher $b_c$ values. Therefore, these results imply that smaller $b_k$ and greater $b_c$ result in greater capacity. The fact that smaller $b_k$ and greater $b_c$ mean that respectively less residual molecules from previous time slots decreases the interference to the molecules absorbed in the current time slot and larger quantity of molecules absorbed within the same time slot as they are transmitted.

%%%%%%%%%%%%%%%%%%%%%%%%%%%%%%%%%%%%%%%%%%%%%%%%%%%%%%%%%%%%%%%%%%
\subsubsection{Number of time slots and time window duration ($n$ and $t_W$)}
\begin{figure}[t!]
\centering
\begin{subfigure}{.75\textwidth}
  \centering
  % include second image
  \includegraphics[width=1\linewidth]{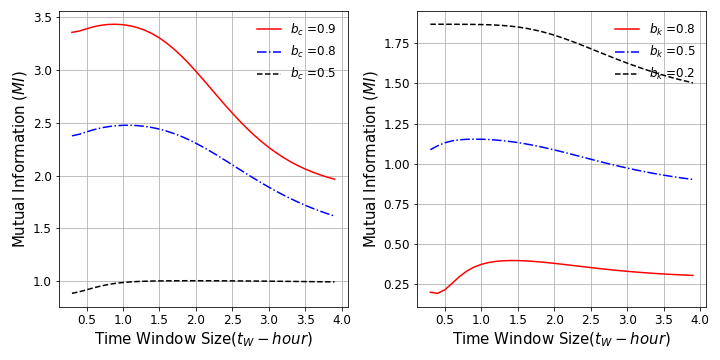} 
  \caption{}
  \label{Res-31}
\end{subfigure}
\begin{subfigure}{.75\textwidth}
  \centering
  % include first image
  \includegraphics[width= 1\linewidth]{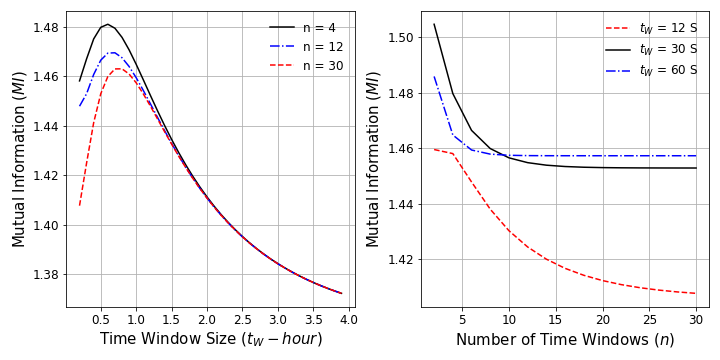}  
  \caption{}
  \label{Res-32}
\end{subfigure}
\caption{ Variability in the mutual information $MI$ with respect to variations of (a) time window size $t_W$, and transmission probabilities $b_c$ and $b_k$, and (b) number of time windows $n$ and time window size $t_W$, given that $Ep= 80$.}
\label{R3}
\vspace{-.75cm}
\end{figure}

Figure \ref{Res-31} shows the variability in the mutual information $MI$ with time widow size (or duration) ($t_W$) for three different transmission probabilities $b_c$ and $b_k$ values when the parameters are set as threshold $Ep = 80 $ and $n= 20$. Although the mutual information $MI$ curves show a slight increase and achieve maximum when $t_W$ is around $1-1.5$ hours, they decrease with increasing time window duration $t_W$. This is based on the fact that the glucose concentration reaches its maximum around within $1-1.5$ hours as depicted in Figure \ref{Res-12}. Also, the variability in the mutual information $MI$ with time window duration $t_W$ for different transmission probabilities $b_c$ and $b_k$  results in similar outcomes as in Figure \ref{R2}. That is, the mutual information $MI$ is higher for the greater $b_c$ and smaller $b_k$ values. When the transmission probabilities $b_c= 0.9, b_k=.0001$ and the threshold $Ep = 80$, Figure \ref{Res-32} shows the impact of $n$ and $t_W$ together on the mutual information $MI$. For smaller $n$, the mutual information $MI$ is  the larger and that is because of larger $n$ results in greater \emph{ISI} noise in the current time slot. Figure \ref{Res-32} also show higher mutual information $MI$ for longer time window duration. This is because the larger time window duration increases the chance of molecules reaching the receiver effectively. 

Moreover, Figure \ref{Res-32} depicts that the mutual information $MI$ increases with time window size $t_W$ and reaches its peak at $0.5-1$ hours. This because both the starch and glucose concentrations increases in the SI tract during this time as shown in Figure \ref{Res-12}, where the channel is filled with higher quantity of starch and glucose molecules and this in turn slows down the propagation speed. As a result, the mutual information $MI$ starts decreasing. In addition, Figure \ref{Res-41} shows the change in the mutual information $MI$ with the threshold $Ep$ for four different combinations of time window size $t_W$ and number of time window $n$ values, given the transmission probabilities are set in the range of $b_c = [0.5-0.9]$ and $b_k = [0.001-0.2]$. With the increase in the threshold $Ep$, the mutual information $MI$ reaches its peak ($\sim 3.5$) and then saturates at a certain value of $EP$, which is higher for shorter $t_W$ and larger $n$. However, when the threshold $Ep \geq 80$, the mutual information $MI$ stays at its maximum value regardless of the changes in $n$ and $t_W$. Therefore, the mutual information $MI$ is greater for larger time window size $t_W$ and smaller  number of time windows $n$.

%%#################################################################
\subsubsection{Impact of velocity, Brownian noise and distance ($u, \sigma$ and $L$)}

\begin{figure}[t!]
\centering
\begin{subfigure}{.35\textwidth}
  \centering
  % include second image
  \includegraphics[width=1\linewidth]{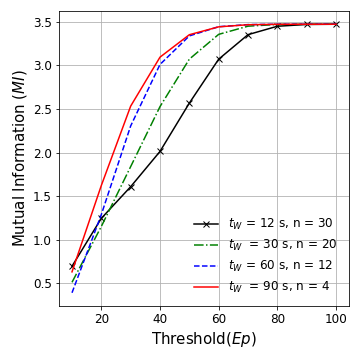}
  \caption{}
  \label{Res-41}
\end{subfigure}
\begin{subfigure}{.35\textwidth}
  \centering
  % include first image
  \includegraphics[width= 1\linewidth]{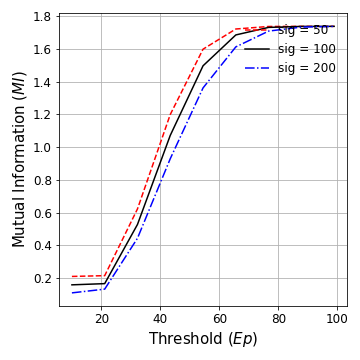}  
  \caption{}
  \label{Res-42}
\end{subfigure}
\begin{subfigure}{.35\textwidth}
  \centering
  % include first image
  \includegraphics[width= 1\linewidth]{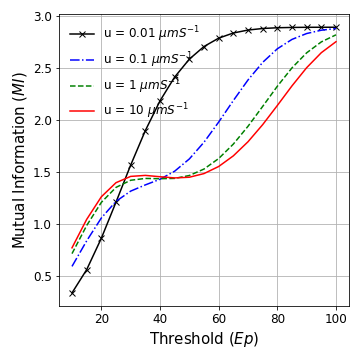}  
  \caption{}
  \label{Res-43}
\end{subfigure}
\begin{subfigure}{.35\textwidth}
  \centering
  % include first image
  \includegraphics[width= 1\linewidth]{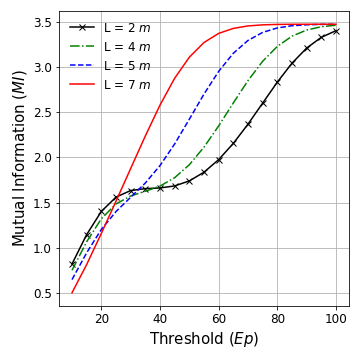}  
  \caption{}
  \label{Res-44}
\end{subfigure}
\caption{Variations and combinations of parameters and its impact on the mutual information $MI$ with respect to variations in the (a) time window size $t_W$ and number of time windows $n$, (b) Brownian noise ($\sigma$ denoted as $sig$), (c) velocity $u$ and (d) molecule traveling distance $L$.}
\label{R4}
\vspace{-.75cm}
\end{figure}

For the same values of the transmission probabilities $b_c$ and $b_k$ used in Figure \ref{Res-41}, Figure \ref{Res-42} shows that the variability in the mutual information $MI$ with threshold $Ep$ for different Brownian noise levels (where $sig = \sigma^2$) when the parameters $t_W$ and $n$ are fixed to $60$ min and $10$, respectively. The mutual information $MI$ increases with the threshold $Ep$ and then saturates after achieving its maximum which is approximately $1.8$ and the mutual information $MI$ is less for larger $sig$ values. This is based on the fact that receiving of glucose molecules results in decreasing the background noise. Considering the influence of velocity of the glucose molecules along the  SI tract on the mutual information $MI$, Figure \ref{Res-43} depicts that mutual information $MI$ is greater for smaller velocity for the same values of number of time slots $n$ and time window size $t$ as used in Figure \ref{Res-42}. The smaller the velocity, the greater the mutual information $MI$ and this is because lower velocity increases the opportunity for the enzymes to react to the starch molecules and increases the glucose production. At the same time the receiver has higher chances of receiving glucose molecules effectively as a result. The distance that the glucose molecules travel along the SI tract also has an impact on the  glucose absorbed through the SI wall because Figure \ref{Res-44} shows that for longer traveling distance, the mutual information $MI$ is higher and also the threshold $Ep$ that maximize the mutual information $MI$ is smaller. These results, therefore, suggest that the mutual information $MI$ reaches to its maximum for smaller $\sigma$ and $u$ and greater time window $t$ and length of the tract $L$. 

All in all, these outcomes suggest that the transmission probabilities with small $b_k$ ($\sim 0$) and larger $b_c$ ($\sim 1$), as well as larger time window size $t_W$ and smaller number of time windows $n$ contribute to maximizing the mutual information $MI$. Moreover, the digestive system related factors such as low digest velocity $u$, small Brownian noise $\sigma$ and longer molecule traveling distance $L$ also make a significant impact on increasing the mutual information $MI$. Therefore, a greater digestive capacity can be observed by keeping these parameters in their optimal ranges as these outcomes are suggested.

%%#################################################################
\subsection{Starch Digestion Capacity }
By using the insights derived through exploring the impact of different parameters on maximizing the mutual information $MI$, this section presents variability in the small intestine capacity $C$ with respect to a set of digestive system parameters which play different key roles to maintain the digestive system functionality at an optimum level.

\begin{figure}[t!]
\centering
\begin{subfigure}{.35\textwidth}
  \centering
  % include second image
  \includegraphics[width=1\linewidth]{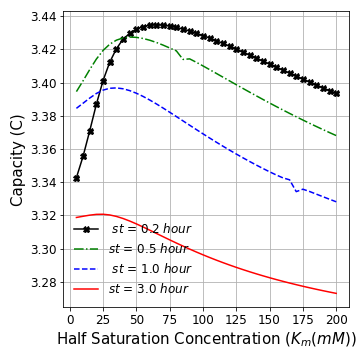}
  \caption{}
  \label{Res-51}
\end{subfigure}
\begin{subfigure}{.35\textwidth}
  \centering
  % include first image
  \includegraphics[width= 1\linewidth]{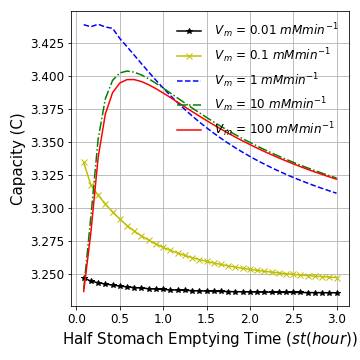} 
  \caption{}
  \label{Res-52}
\end{subfigure}
\begin{subfigure}{.35\textwidth}
  \centering
  % include first image
  \includegraphics[width= 1\linewidth]{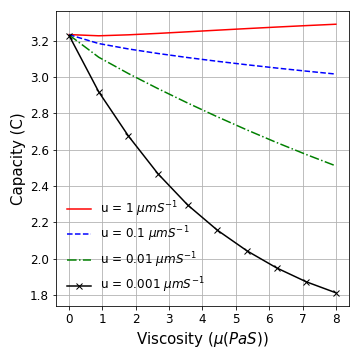}  
  \caption{}
  \label{Res-53}
\end{subfigure}
\begin{subfigure}{.35\textwidth}
  \centering
  % include first image
  \includegraphics[width= 1\linewidth]{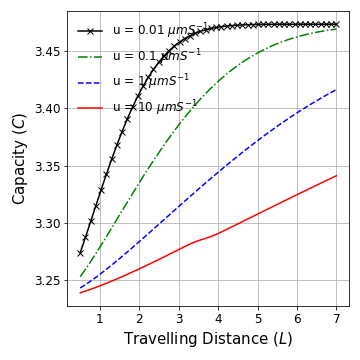}  
  \caption{}
  \label{Res-54}
\end{subfigure}
\caption{Variability in capacity with the change in combination of parameters (a) half stomach emptying time ($st$) and half saturation concentration of starch ($K_m$), (b) maximum reaction rate of starch with enzymes ($V_m$) and half stomach emptying time ($st$), (c) velocity ($u$) and viscosity ($\mu$) of starch (d) velocity ($u$) and molecule traveling distance $L$.}
\label{R5}
\vspace{-.75cm}
\end{figure}

Figure \ref{R5} shows the small intestine capacity $C$ with respect to changes in the half-stomach emptying time $st$, half saturation concentration $K_m$ of the maximum reaction rate $V_m$, viscosity $\mu$, velocity $u$ and traveling distance $L$ when the remaining parameters are set as $n = 10$, $t= 60$ min, $b_c = [0.5-0.9]$, and $b_k = [0.001-0.2]$. Figure \ref{Res-51} shows the change in the capacity with half saturation concentration $K_m$ for four different half-stomach emptying time $st$ values. The capacity reaches its peak at some point of $K_m$ which is around $25-75 mM$ for the selected half-stomach emptying time $st$ values. This is because when the half saturation concentration $K_m$ is in this range, the abundance of enzymes increases the reaction on the starch and the production of glucose as a result. The decline in the small intestine channel capacity with further increase in the half saturation concentration $K_m$ means that the starch concentration required to reach the half of the maximum reaction rate is high and this results in slow glucose production speed.  Also, the longer the half-stomach emptying time $st$, the smaller the small intestine channel capacity. That is because the longer $st$ decreases the gastric emptying rate $\gamma (= \log(2)/st)$ and hence the digest product enters into the SI tract slowly which in turn results in low glucose production rate. Moreover, the glucose absorbance capacity  also depends on the maximum reaction rate $V_m$ as depicted in Figure \ref{Res-52}, which shows  the change in capacity $C$ with half gastric emptying time $st$ for five different maximum reaction rate values. The channel capacity decreases with increasing half gastric emptying time $st$ when $V_m \leq 1 (mMmin^{-1})$ and for maximum reaction rate $V_m > 1 (mMmin^{-1})$, the channel capacity $C$ increases with  half-stomach  emptying  time $st$ to approximately $0.75-1$ hour and then decreases. The highest channel capacity $C$ is observed when maximum reaction rate $V_m= 1 mMmin^{-1}$. Thus, this suggests that the maximum reaction rate $V_m \geq 1$ results in a reasonable starch digestion and glucose absorption capacity. The maximum reaction rate could, however, be varied with other parameters such as food viscosity, velocity and traveling distance. 

Figure \ref{Res-53} shows how the capacity changes when the viscosity $\mu$ changes over the range  $[0.001-8]PaS $ for four different velocity values. The higher the viscosity and the velocity, the smaller the capacity. However, when the velocity $u$ is increasing, the rate at which the channel capacity $C$ declines with $\mu$ becomes smaller and as a result the channel capacity $C$ shows an increasing trend when the velocity $u = 1ms^{-1}$, where the channel capacity $C$ increases with $\mu$. This is because the high viscosity reduces the chance of enzyme reactions on the digest product and also the velocity of the  molecules and the production of glucose. Besides, the glucose traveling distance over the SI tract seems to have a significant impact on the channel capacity $C$ as shown in Figure \ref{Res-44}, which shows the impact of the length $L$ and velocity $u$ on the channel capacity $C$. The longer the glucose traveling distance, the greater the channel capacity. This is based on the fact that the longer traveling distance increases the chance of the digest product to react with the enzymes for the production and absorption of glucose. Also, when the velocity is small, it allows more time for the SI wall to absorb nutrients effectively which in turn increases the channel capacity. For this reason Figure \ref{Res-44} depicts greater channel capacity for smaller velocity and shorter traveling distance.

%%#################################################################
\section{Discussion}\label{sec-4}
The digestive capacity model proposed here helps in exploring the influence of both the factors associated with the digestive system and personal physiological characteristics that have a role in the digestive system dynamics. For instance, Figure \ref{Res-51} and \ref{Res-52} depicts the impact of gastric emptying time on the glucose absorbance capacity. This variability could explain the influence of physiological factors associated with the stomach such as aging and low activity level, which results in longer gastric emptying time and increase in the glucose absorbance delay. Also, Figure \ref{R4} and \ref{R5} show that different interference due to velocity turbulence and viscosity of digest generated in the digestive system reduces the small intestine channel capacity. Thus, given an individual personal data, the model can help in understanding the influence of  different factors on the digestive dynamics, thereby providing more personalized insights that may have the potential in customizing food consumption.    

%The study primarily considered carbohydrates digestion dynamics as it can help in revealing valuable information that can be useful for understanding digestion abnormalities. 
This paper assumed that the carbohydrate digestion happens only in the SI tract. That is because only around $5\%$ of the consumed carbohydrate is digested in the mouth, but the chemical digestion of carbohydrate in the stomach is low due to low pH value. However, carbohydrate may not fully digested and leave the SI due to the inadequacy of enzymes, which is related to the pancreatic inefficiency, resulting low digestive capacity. The half-saturation concentration $K_m$ is associated with the abundance of the enzyme and smaller $K_m$ implies that a greater glucose production (i.e., sufficient enzymes to react on starch). Figure \ref{Res-51} shows that the capacity is higher for smaller $K_m$ which means that the abundance of enzymes in the SI tract helps in the effective conversion of starch into absorbable glucose. Thus, investigating any deviation in this behavior may give an early warning regarding any abnormality in the digestion process. 

The digestion capacity can also be varied due to the lack of brush broader enzymes associated with decreased villus length, and this can result in celiac disease for example \cite{43}. Figure \ref{Res-51} and \ref{Res-54} shows the impact of the enzymes and molecules traveling length on the digestion capacity. Similarly, the impact of the velocity and viscosity on the channel capacity depicted in Figure \ref{Res-43} and \ref{Res-53} can explain the malabsorption due to the increased undigested portion of starch that leaves the SI tract. This is because the high viscosity limits enzyme activities on the starch and the high velocity does not provide sufficient time for the enzymes to react on the  starch, resulting in constipation and diarrhea. Therefore, the capability in characterizing such problems due to the efficiency of the digestive system can be helpful in designing effective treatments.

%%#################################################################
\section{Conclusion}\label{sec-5}
This paper gives a MC representation for the digestive system and then proposes an advection-diffusion and reaction mechanisms based model to characterize the digestive capacity in the small intestine. The influence of different physiological factors related to the digestive system and consumed food on the channel capacity dynamics of the digestive system is explored in the context of carbohydrates digestion. The numerical results show that the shorter gastric emptying time and low half-saturation concentration, small velocity and viscosity, and longer traveling distances of the digest product (i.e., starch) increases the digestive capacity. Also, the digestive capacity is highest when there is low interference which mainly occurs due to the delay in propagation of glucose and starch molecules. These insights can increase the potential of understanding and characterizing different digestive system dynamics such as abnormalities in food consumption patterns and digestion efficiency that depends on different personal physiological settings and internal digestion conditions, in order to provide more personalized recommendations. Our proposed approach provides a new communication-theoretic tool for characterizing the functionalities of the digestive system, providing a new analysis tool to understanding the impact of food types and their propagation behavior within the SI tract. By analyzing this from the communication channel capacity, and provide a new mechanism of measuring efficiency in the digestive process.

%they can be used for providing more personalized recommendations depending on different personal physiological settings and internal digestion conditions. Moreover, they can also contribute to accelerating the development of a digital twin for the digestive system which would enable integrating the digestive system with modern communication technologies to explore the digestive dynamics more in-depth in the future. %This in turn can contribute to minimizing global food-related such as obesity, food waste and sustainability.    

%%#################################################################
\section*{Appendix}
\subsection*{Food Transport Model}
\begin{figure}
    \centering
    %\vspace{-2.0cm}
    \includegraphics[width = .45\textwidth]{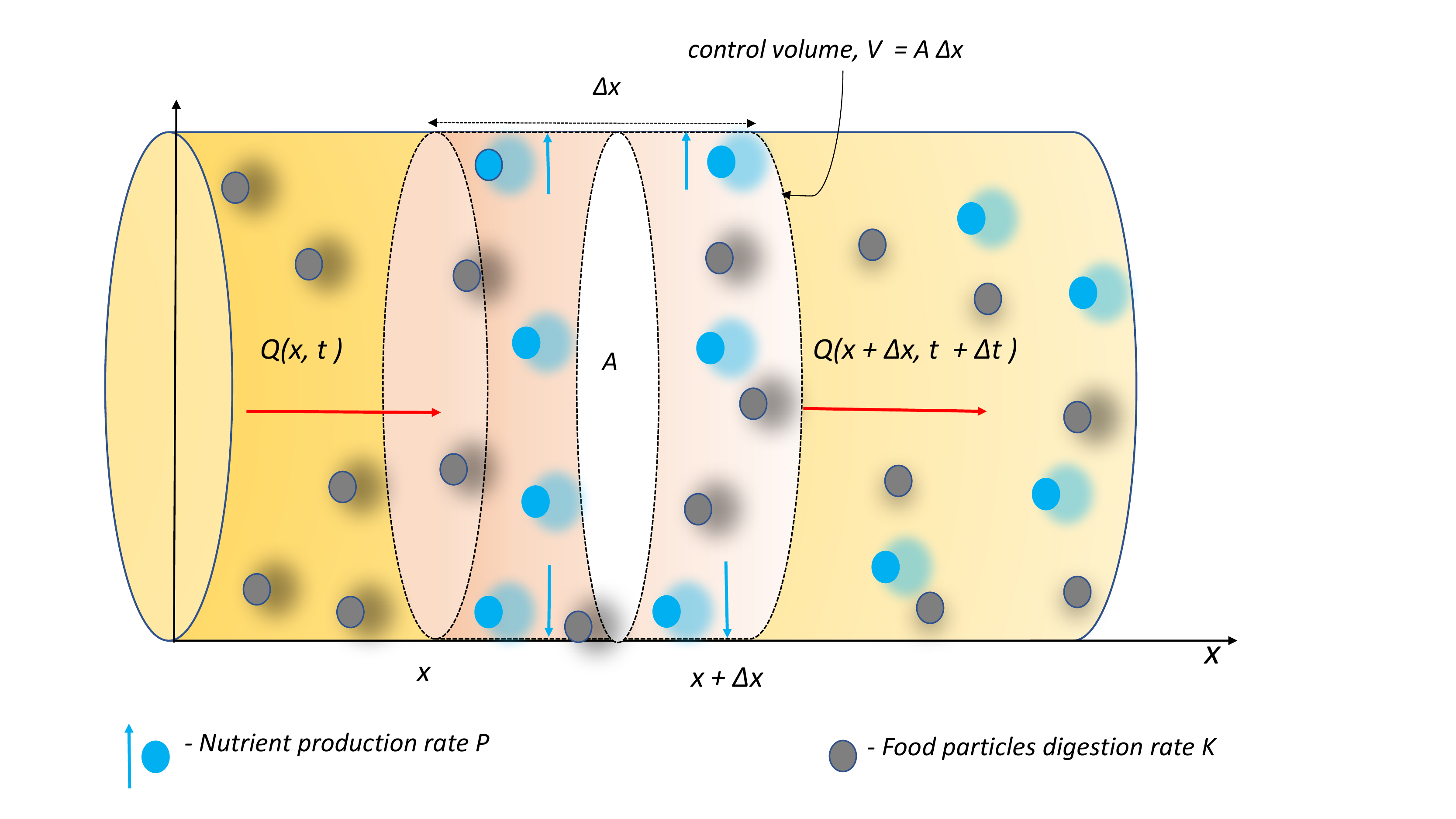}
    %\vspace{-1.5cm}
    \caption{Propagation and digestion of food particles into absorbable nutrients.}
    \label{Transport}
\end{figure}

Figure \ref{Transport} illustrates a propagation of food particles through a control volume $V$ ($= A\delta x$, where $A$ is the cross sectional area of the control volume) of depth $\delta x$ in the SI tract with a mass flow rate $Q (gm^{-2}s^{-1})$. The rate of change of the food mass through the control volume can be expressed in the following mass balance expression

\begin{equation}\label{nu-0}
  V\dv{C(x,t)}{t} = \textrm{Mass}_{in} -  \textrm{Mass }_{out},
\end{equation}
where the terms, inlet and outlet mass of food ($\textrm{Mass}_{in}$ and $\textrm{Mass}_{out}$) in  (\ref{nu-0}) are formulated as 
\begin{eqnarray}\label{nu-1}
    \textrm{Mass}_{in} &=& Q(x,t)_{in} A + \textrm{Conversion of food into nutrients}, \nonumber \\
    &=& Q(x,t)_{in} A + C_P(x,t)V, \label{mass-1} \\
    \textrm{Mass}_{out} &=& Q(x+\delta x,t)_{out} A + \textrm{Absorption of nutrients}, \nonumber \\
    &=& Q(x + \delta,t)_{out} A + K_aC(x,t)M, \label{mass-2}
\end{eqnarray}
where $C(x,t) (gm^{-3})$ is the nutrient concentration and $C_P(x,t) (gm^{-3}s^{-1})$ and $K_a (ms^{-1})$ are the nutrient production (or reaction) and absorption rates, respectively and $M = (2\pi d \delta x) f$ is the surface area of the controlled volume ($d$ is the radius of the SI tract and $f$ is the surface area with increased factors due to villus folds in the SI wall). By substituting (\ref{mass-1}) and (\ref{mass-2}) in (\ref{nu-0}),
the change in concentration can be expressed as   
\begin{equation}\label{nu-2}
    \frac{\partial C(x,t)}{\partial t} = -\frac{\partial Q(x,t)}{\partial x} - \frac{2}{d}K_af C(x,t) + C_P(x,t).
\end{equation}

Considering the advection-diffusion based flow of food stuff in the SI tract, the mass flow rate of food stuff can be written by using the Fick's first law as 
\begin{equation}\label{nu-3}
    Q(x,t) = C(x,t)u - D \frac{\partial C(x,t)}{\partial x},
\end{equation}
where $u (ms^{-1})$ is the average velocity of the food particle flow in the SI tract. Then, the partial derivative of (\ref{nu-3}) with respect to traveling distance, $x$, can be expressed as in (\ref{nu-4}) and represented as follows 

\begin{equation}\label{nu-4}
    \frac{\partial Q(x,t)}{\partial x} = u\frac{\partial C(x,t)}{\partial x} - D \frac{\partial^2 C(x,t)}{\partial x^2}. 
\end{equation}

Combining (\ref{nu-2}) and (\ref{nu-4}), the change in nutrient concentration can be expressed as in (\ref{App-nu-6}).  In general, (\ref{App-nu-6}) is known as the governing equation of the advection-diffusion-reaction based on fluid flow in a pipe and represented as 

\begin{equation}\label{App-nu-6}
    \frac{\partial C(x,t)}{\partial t} + u \frac{\partial C(x,t)}{\partial x} = D \frac{\partial^2 C(x,t)}{\partial x^2} - K C(x,t) + C_P(x,t),  
\end{equation}
where $K = \frac{2}{d}K_af$.
%%#################################################################

\subsection*{Nutrient Production and Absorbance Rates ($P$ and $K$)}
\begin{itemize}
    \item {\bf Nutrient production rate ($C_P(x,t)$)} can be expressed by the chemical expression given in (\ref{s-g-rection}). This reaction represents the enzyme ($E$) reaction on food stuff ($S$) (e.g., starch) that produces an absorbable nutrient ($C$) (e.g., glucose) represented by the chemical reaction as follows 
    \begin{equation}\label{s-g-rection}
        E + S \underset{k_d}{\stackrel{k_a}{\rightleftharpoons}} ES {\stackrel{k_p}{\rightarrow}} C + E.
    \end{equation}
    Based on the Michaelis-Menten-kinetics \cite{18}, the nutrient production rate ($C_{P(x,t)}$) can be expressed as 

    \begin{equation} \label{nu-p}
         C_{P(x,t)} = V_{m} \frac{[S]}{K_{m} + [S]},
    \end{equation}
    where $V_m$ is the maximum reaction rate achieved at the maximum saturating concentration and the half saturation concentration $K_{max} = \frac{k_p + k_d}{k_a}(Ms^{-1})$. Here $k_a (Ms^{-1}), k_d (s^{-1})$ and $k_p(s^{-1})$ are association, disassociation and nutrient production rates. 

    \item {\bf The absorption rate ($K_a$)} is computed by using the relationship between the Sherwood number ($Sh$), Reynolds number ($Re$) and Schmidt number ($Sc$) given in \cite{18} and represented as follows 
    
    \begin{equation}
       Sh = 1.62 Re^{\frac{1}{3}} Sc^{\frac{1}{3}} (d/L)^{\frac{1}{3}},
   \end{equation}
    where $Sh = K_a \frac{d}{D}$, $Re = \frac{\rho ud}{\mu}$, and $Sc = \frac{\mu}{\rho D}$. Here $\rho$ and $\mu$ are the fluid density and viscosity, respectively. 
   
   Then the mass transfer coefficient, $K_a$, is computed as 
   \begin{equation}\label{nu-ab-s}
       K_a = 1.62 \left(\frac{u D^2}{Ld}\right)^{\frac{1}{3}},
   \end{equation}
    where diffusion $D = \frac{K_BT}{6\pi \mu r_m} (m^2 s^{-1})$, $K_B (m^2kg s^{-1}K^{-1})$ is the Boltzmann constant, $T (K)$ represents the absolute temperature and $r_m (nm)$ is the radius of the diffusing nutrient molecules and $L (m)$ and $d (cm)$ are the length and diameter of the SI tract, respectively. 
   \end{itemize}
%%#################################################################
\subsection*{Analytical Solution}
Analytical solutions for (\ref{eq-1})-(\ref{eq-3}) are derived by assuming $K_m >> C_s$. Here we present the derivation for the glucose derivation $C_g(x,t)$ with respect on distance $x$ and time $t$.
\begin{itemize}
    \item [(1)] {\bf Concentration of Starch in the Stomach:} Suppose at $t=0$, $C_0$ is the amount of starch consumed. The solution of (\ref{eq-1}) represents the variability in starch concentration in stomach ($C_{st}$) over time $t$ and is represented as   
    \begin{equation} \label{eq-1-sol}
        C_{st}(t) = C_0 e^{-\gamma t}.
    \end{equation}
    
    \item [(2)] {\bf Concentration of Starch in the SI Tract:} When $K_m >> C_s(x,t)$, $ \displaystyle V_m\frac{C_s}{K_m + C_s} \approx \frac{V_m}{K_m}C_s$. 

    Let $C_s(x,t) = f(x,t)e^{\alpha x - \beta t} + pe^{-\gamma t}$ and substitute this in (\ref{eq-2}). When $ \displaystyle \alpha = u/2D_s, \beta = u^2/4D_s + V_m/K_m$ and $\displaystyle p = \frac{\gamma C_0 K_m}{V_m - \gamma K_m}$, (\ref{eq-2}) can be simplified into a pure diffusion process as given in (\ref{diff}) along with its analytical solution ($f(x,t)$)\cite{3} and is represented as follows
    
    \begin{equation}\label{diff}
        \frac{\partial f(x,t)}{\partial t} = D_s \frac{\partial^2 f(x,t)}{\partial x^2}   \rightarrow   f(x,t) = \frac{f(0,0)}{\sqrt{2 \pi D_s t}} exp\left(-\frac{x^2}{2Dt}\right).
    \end{equation}
    By substituting the value of $f(x,t)$ given in (\ref{diff}) in $C_s(x,t)$, the analytical solution of (\ref{eq-2}) is given by (\ref{eq-2-sol}), given that  $C_s(x=0,t=0) = 0$ and is represented as
        \begin{equation} \label{eq-2-sol}
            C_s(x,t) = G( e^{-\gamma t} -Q),
        \end{equation}
     where $\displaystyle  G = \frac{\gamma C_0 K_m}{ V_m - \gamma K_m- }$ and $ \displaystyle Q =   \frac{1}{\sqrt{4\pi D_st}}e^{\left(\frac{-(x-ut)^2}{4D_st} - \frac{V_m}{K_m}t \right)}$.   
    
    \item [(3)] {\bf Concentration of Glucose in the SI tract:} Following the similar approach as in step 2, we substitute the expression given in (\ref{subs-1}) into (\ref{eq-3}) and this is represented as 
    \begin{equation}\label{subs-1}
    C_g(x,t) = f(x,t)e^{\alpha x - \beta t} + qC_s(x,t).
    \end{equation}
    (\ref{eq-3}) can be simplified  into  (\ref{diff}), when $\alpha = u/2D_g, \beta = u^2/4D_g +K$ and  $q = \frac{V_mC_s}{K_m(C_{st} + uC_{sx} - D_gC_{sx^2}+KC_s)}$, where $C_{st} = \dv{C_s}{t}, C_{sx} = \dv{C_s}{x}$ and $C_{sx^2} = \dv{^{2}C_s}{x^2}$. By assigning $C_s, C_{st}, C_{sx}$ and $C_{sx^2}$ in $q$, $q$ can further be simplified as $q = \frac{V_m(Q- e^{-\gamma t})}{( K-\gamma )K_m e^{-\gamma t} + Q(V_m -KK_m)}$.
    
    When $x=0$ and $t = 0$,  $f(x,t)$ can be expressed as follows 
    \begin{equation}
         f(0,0) = -q C_s(0,0) =  \frac{C_0 \gamma V_m}{( K-\gamma )(V_m -\gamma K_m)}.
    \end{equation}
Therefore, the solution of (\ref{eq-3}) can be written as 
    \begin{equation}\label{eq-3-sol}
        %C_g(x,t) = \frac{C_0\gamma V_m}{(V_m - KK_m)(K- \gamma)}\left[ \frac{1}{\sqrt{4\pi D_g t}} e^{\frac{-(x-ut)^2}{4D_g t} }\left( e^{-Kt}  - e^{-\frac{V_m}{K_m}t}\right) + e^{-\gamma t} \right].
        C_g(x,t) = \frac{C_0\gamma V_m}{(V_m - KK_m)(K- \gamma)}\left[ \frac{1}{\sqrt{4\pi D_g t}} e^{\left( \frac{-(x-ut)^2}{4D_g t} -Kt \right)} - \frac{1}{\sqrt{4\pi D_s t}} e^{\left (\frac{-(x-ut)^2}{4D_s t} -\frac{V_m}{K_m}t \right) }
      + e^{-\gamma t} \right].
    \end{equation}
\end{itemize}
Finally, the glucose concentration along the SI tract is computed  by integrating (\ref{eq-3-sol}) over the SI tract ($L$) and is represented as follows

%\begin{eqnarray}\label{eq-3-sol-1}
    %C_{G}(t) &=&  \int_0^{d} C_g(x,t)dx \\
    %&=& \frac{C_0 \gamma V_m}{(V_m - KK_m)(K-\gamma)} \left[ \frac{d}{2}\left( e^{-Kt} - e^{-\frac{V_m}{K_m}t}\right)  \left( erf \left(\frac{d-ut}{2\sqrt{Dt}}\right) + erf \left(\frac{ut}{2\sqrt{Dt}}\right) \right)+ de^{-\gamma t} \right]
   % C_{G}(t) &=&  \int_0^{d} C_g(x,t)dx \\
    %&=& \frac{C_0 \gamma V_m}{(V_m - KK_m)(K-\gamma)} \left[ \frac{d}{2} e^{-Kt} \left( erf \left(\frac{d-ut}{2\sqrt{D_gt}}\right) + erf \left(\frac{ut}{2\sqrt{D_gt}}\right) \right) - 
    %\frac{d}{2}e^{-\frac{V_m}{K_m}t} \left( erf \left(\frac{d-ut}{2\sqrt{D_gt}}\right) + erf \left(\frac{ut}{2\sqrt{D_gt}}\right) \right) +
    %de^{-\gamma t} \right]
%\end{eqnarray}

\begin{multline*}\label{eq-3-sol-1}
     C_{G}(t) = \frac{C_0 \gamma V_m}{(V_m - KK_m)(K-\gamma)} \left[ \frac{L}{2} e^{-Kt} \left( erf \left(\frac{L-ut}{2\sqrt{D_gt}}\right) + erf \left(\frac{ut}{2\sqrt{D_gt}}\right) \right)\right] \\
    \frac{C_0 \gamma V_m}{(V_m - KK_m)(K-\gamma)} \left[ \frac{L}{2}e^{-\frac{V_m}{K_m}t} \left( erf \left(\frac{L-ut}{2\sqrt{D_gt}}\right) + erf \left(\frac{ut}{2\sqrt{D_gt}}\right) \right) +
    de^{-\gamma t} \right].
\end{multline*}

\section*{Acknowledgement}
This research was supported by a research grant from Science Foundation Ireland and the Department of Agriculture, Food and Marine on behalf of the Government of Ireland under the Grant 16/RC/3835 (VistaMilk).

\bibliographystyle{unsrt}
\bibliography{Paper}

\end{document}